\documentclass[12pt]{article}
\usepackage{graphicx}
\begin{document}
\centerline {WHY EVERYTHING GETS SLOWER ?}   
\bigskip

\centerline {D.Stauffer$^{a,b}$ and K.Ku{\l}akowski$^a$}

\bigskip
\centerline{$^a$ Faculty of Physics \& Nuclear Techniques, University of Mining \& Metallurgy} 

\centerline{al.Mickiewicza 30, 30-059 Krak\'ow, Poland}
\bigskip

\centerline {$^b$ visiting from Inst. for Theor. Physics, Cologne 
Univ., D-50923 K\"oln, Euroland}

\begin{abstract}

A social system is represented by the Barab\'asi-Albert model. At each node of
the graph, an Ising spin is placed, $S=\pm 1$, with antiferromagnetic
interaction between connected nodes. The time to reach equilibrium via
Glauber kinetics 
does not depend on the system size. The average energy associated with the 
rare spin flips in equilibrium
oscillates with the number $m$ of edges of new nodes.
The conclusions are illustrated with events from recent European history,
where after some strong change a rather immobile society evolved.

\end{abstract} 

\bigskip

Some of us wonder sometimes why it is so difficult to reorganize a social
system \cite{galam}. There are many examples in history when a society
prevents changes just by means of inertia. A commonly accepted example is
provided by the history of Soviet Union in 60's, when reforms of Khrushchev
failed despite apparent lack of objections. In fact, the famous 'perestrojka'
in 80's could happen only because members of the Politburo in Moscow were
pretty sure that Gorbachev does not talk seriously \cite{wl}.  

Here we are interested in the time dependence of a flexibility of a social
system, described in possibly simplest way. We profit from recent works of
Barab\'asi and Albert \cite{basci,ba1}, where social connections are modeled
as edges of graphs. A special algorithm was introduced in Ref.\cite{basci} to
construct the so-called scale-free networks of nodes connected by edges. The
resulting systems were proven to show qualitative similarities with networks
known from different branches of knowledge, as biology, ecology or
linguistics. For a review of this new subject see Ref.\cite{ba1}.

A social system is represented by a scale-free Barab\'asi-Albert network, with
the number $m$ of edges of successively added nodes as the only parameter. 
Ising spins $S=\pm 1$ are assigned to all nodes. The interaction is chosen to
be antiferromagnetic and it is limited to the nodes directly connected by
edges. This choice can be motivated heuristically as follows. There is a
natural tendency of social units (people, enterprises or political fractions)
to differ and to compete. As a result of the competition, one unit of each two
becomes more self-concentrated, egoistic and powerful than the other - and it
wins. In this sense, a stable state of each pair of units is: one up
(positive, subordinated) and one down (negative, prevailing). At the initial
state of a newly created system, however, all units are idealistic and claim to
serve to the community. A stable state of the whole system is attained by a
gradual evolution of the units, which is composed from local competitions
between members of interacting pairs. A similar splitting of an evolving
society into performers of two opposite strategies was described by
Mitchell et al. in terms of rules of cellular automata \cite{mcg}.

Our time evolution accords with the Monte-Carlo Glauber dynamics
\cite{heer}. However we believe that in our problem, the character of the
evolution rule is of secondary importance. The relevant feature is the existence
of the working function - energy in the Ising model - which is minimized apart
from thermal fluctuations. As it was proven by Derrida \cite{der}, in this
case a local equilibrium state is well defined, and the length of the limit
cycle is at most 2. In fact, we need only the existence of an absorbing state,
which is stable against existing disturbances. This condition seems to be
fulfilled also in the world of social phenomena, at least until a Great
Disturbance appears.

In contrast to spin glass simulations since a quarter of
a century, we do not discuss
here phase transitions, ordering below the phase transition temperatures, or
ground states. Instead we look at the initial nonequilibrium dynamics as
a model for the development of social rigidity.

We used the program employed already by Aleksiejuk et al \cite{aa} and replaced
ferromagnetic by antiferromagnetic couplings. The Barab\'asi-Albert network
started with $m$ mutually connected sites and then was grown to $N+m$ sites
according to the standard rule that a newly added site selects $m$ neighbours
from the already existing network sites, with a selection probability
proportional to the number $k$ of neighbours the selected site had before. The
Ising spins, initially all up, where flipped with heat bath (Glauber) dynamics,
i.e. up with probability $x/(1+x)$  and down with probability $1/(1+x)$,
where $x = \exp(-nJ/k_BT)$ is the Boltzmann thermal probability and $n =
\sum_j S_j$ is the sum over all $k(i)$ neighbours of spin $i$, with
$S_j = \pm 1$ and a positive exchange energy $J$. One iteration $t \rightarrow
t+1$ is one Monte Carlo step per spin. During each iteration we counted the
total number of spins which are flipped, the total energy change ($|n|$ summed
over all spin flips), and the magnetization $M = \sum_i S_i$.

Fig.1 for $m=5$ shows that the magnetization decays weakly with time, the total
energy changes strongly, while the number of flipped spins is mostly in
between. Changing the network size $N$ from 2 million to 20,000 gives
similar results, Fig.2. Figs.3 and 4 show a slightly faster equilibration
with $m=1$ or 2, instead of 5, and Fig.5 summarizes this trend for $m=2$, 4, 6, 
8, 10. 

In Fig.2 we see that for $m=5$ the energy change stabilizes in about 100
timesteps, and this characteristic time does not depend on the system size.
(It alse remains about the same if we use lower temperatures.) 
The stable value is a small fraction of the number of nodes, i.e. about $0.005
N$. Similarly, the number of spin flips stabilizes near $0.015 N$. The
magnetization changes reveal two stages; after a short transient, it decreases
with time $t$ proportionally to $t^{-\beta}$, with $\beta $ less than 0.1.

The $m$ dependence of the results, Fig.6, reveals some kind of oscillations with
amplitude decreasing with $m$. This effect can be assigned to the difference
between $m$ even and odd, which is expected to alter the density of
frustration in the system: A cycle with an even number of antiferromagnetic 
edges allows all spins to be antiparallel to their neighbours on this cycle; 
for an odd number of edges this is impossible and leads to frustrated edges
connecting parallel spins even at zero temperature.

In the above results, the flexibility of the modeled system is measured by the
number of spins flipped at a given iteration step, and by the energy change
during these flips. Continuing our allegorical interpretation, we assign the
number of personnel changes to the number of spin flips. This is justified
by our every-day experience that it is easier to change a person than his/her
attitude. Even more important is the change of energy, which can be seen as a
measure of a social cost of changes. An exchange of a politician without a
resulting change in the problems of the society corresponds to a spin flip 
with zero energy change, $n=0$. These flips dominate at long times in our
simulations. 

For social systems, two conclusions can be drawn from our results. First, after
a big (external) disturbance, the
ability of a system to change decreases with time and soon it becomes
very small. Actually, this result follows as a simple consequence of our model
of a social system as an interacting medium with the energy as a work function
which is minimized during the time evolution. With this assumption, the proof
of Derrida \cite{der} holds. Our second conclusion is more subtle. If the above
interpretation is valid, personnel changes or even changes of attitude of
the administration are not equivalent to changes of the social structure,
which can remain unharmed. A real change demands a new initial state far from
equilibrium, when most units agree.

These conclusions can be easily illustrated by examples from recent history of
Europe. In years 1989-91 the reforms of the democratic Polish government
transformed the life in the country much more than the changes in the ten
following years. The characteristic time can then be evaluated as one or two
years. The moral is: once you can change something, do it quickly. On the other
hand, there is a continuity of the so-called 'national interests' of many
countries, and they are present in political strategies through centuries. At
the reverse of the same effect we find that the amounts of unemployed people in
several countries (including ours) seem to be very resistant with respect to
changes of governments and parliaments, not to mention a "New Politics"
claimed frequently during elections.

\begin{figure}[hbt]
\begin{center}
\includegraphics[angle=-90,scale=0.35]{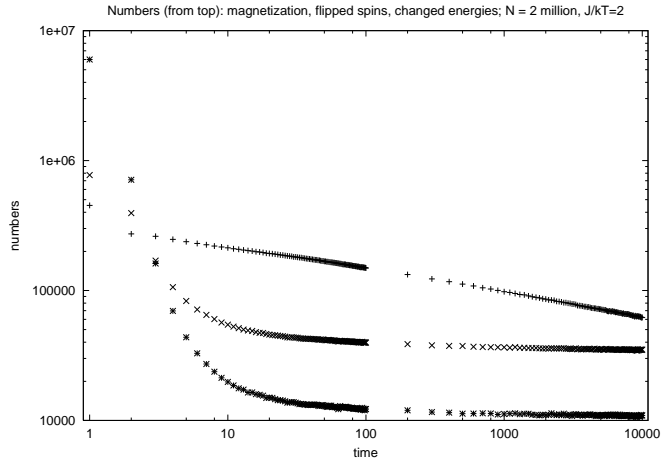}
\end{center}
\caption{Relaxation for $m=5$ at $k_BT/J$ = 0.5 for two million nodes.
}
\end{figure}

\begin{figure}[hbt]
\begin{center}
\includegraphics[angle=-90,scale=0.35]{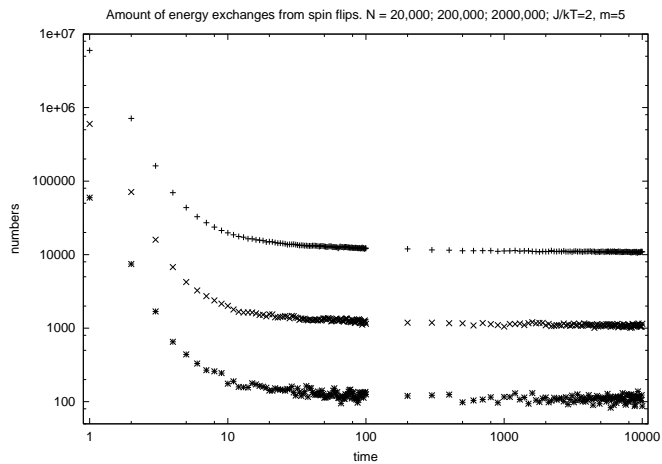}
\end{center}
\caption{Sum of energy changes for small, medium and large systems; again
$m=5$ at $k_BT/J$ = 0.5.
}
\end{figure}

\begin{figure}[hbt]
\begin{center}
\includegraphics[angle=-90,scale=0.35]{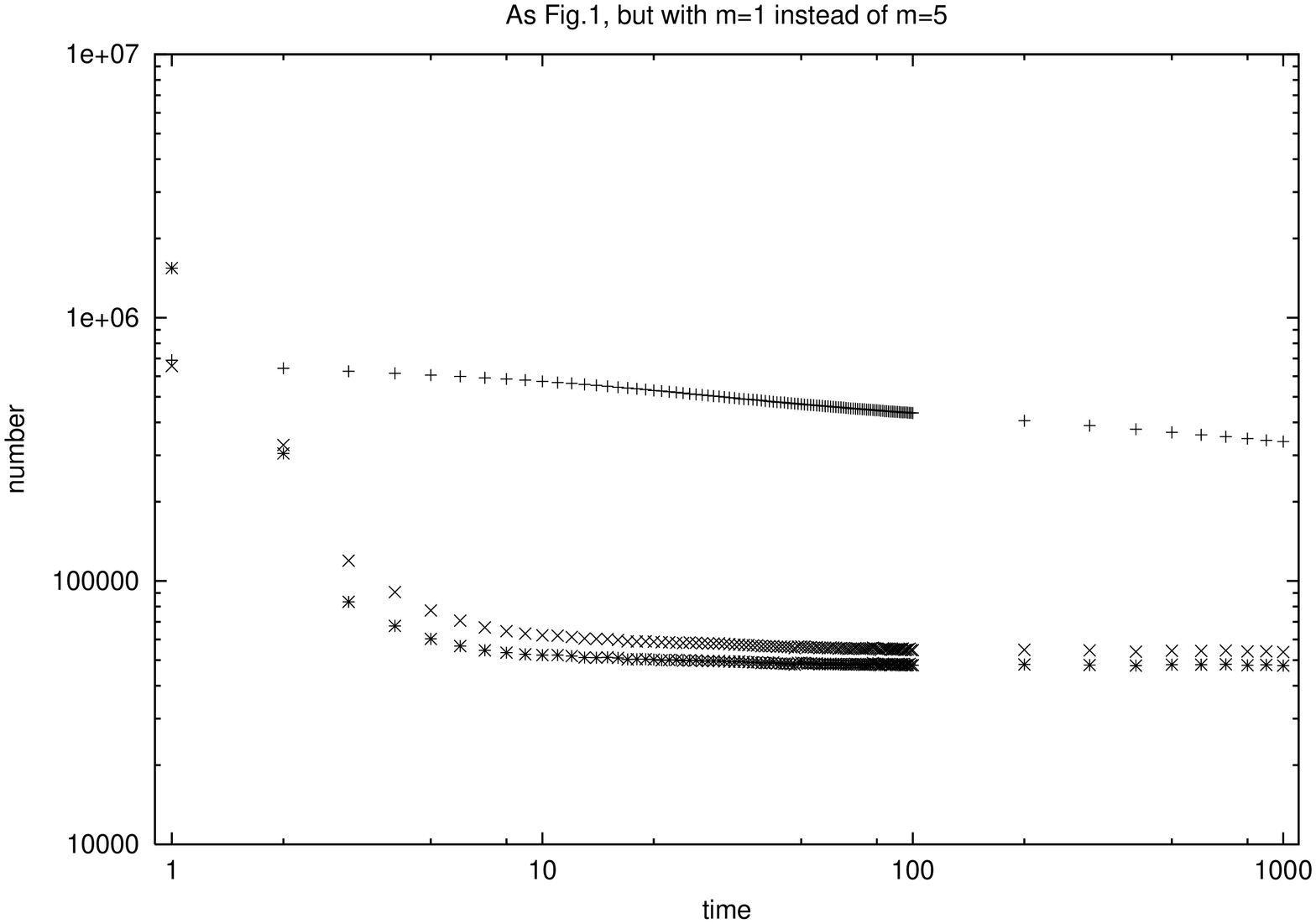}
\end{center}
\caption{As Fig,1 but with $m=1$ instead of $m=5$.
}
\end{figure}

\begin{figure}[hbt]
\begin{center}
\includegraphics[angle=-90,scale=0.35]{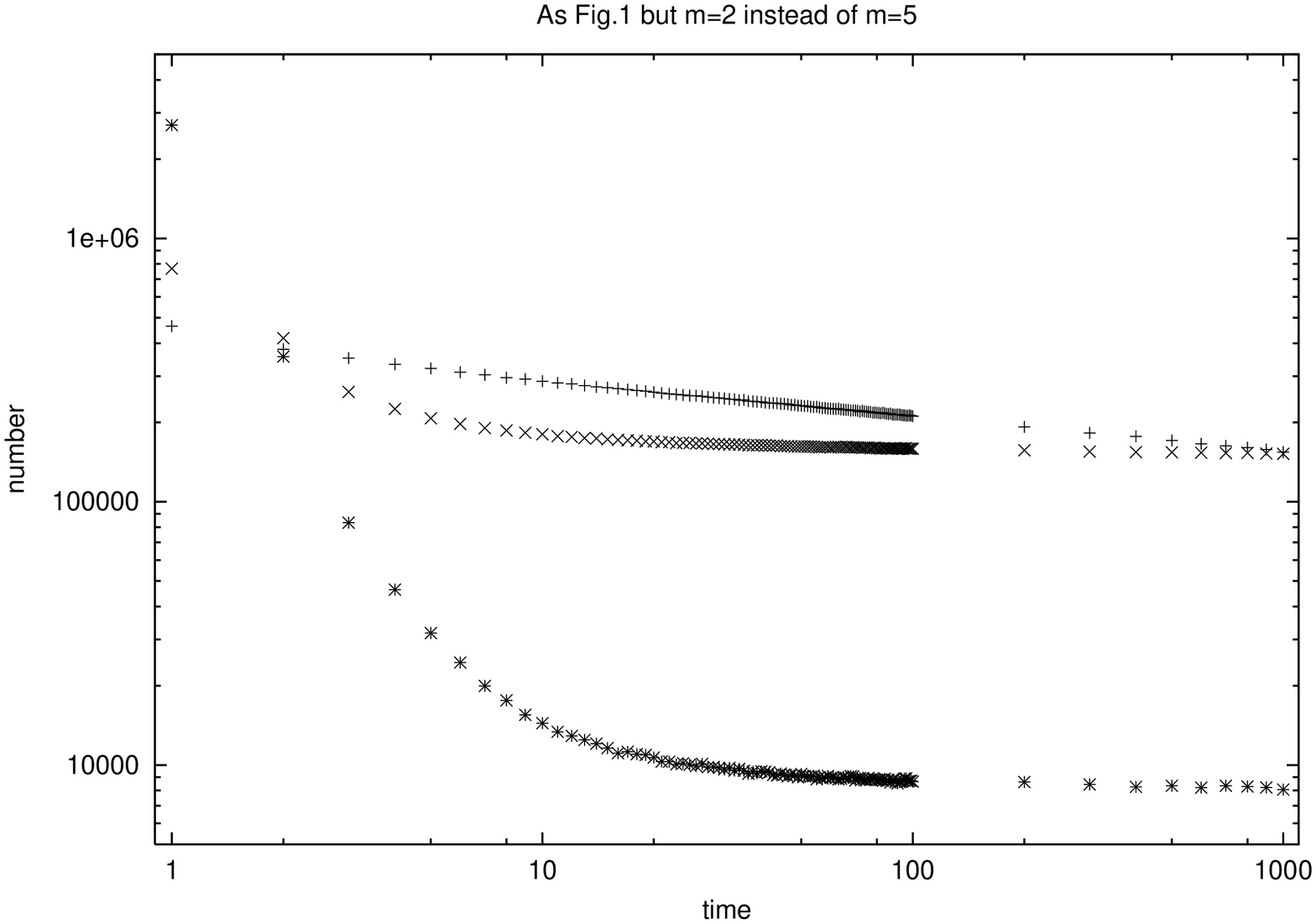}
\end{center}
\caption{As Fig,1 but with $m=2$ instead of $m=5$.
}
\end{figure}

\begin{figure}[hbt]
\begin{center}
\includegraphics[angle=-90,scale=0.35]{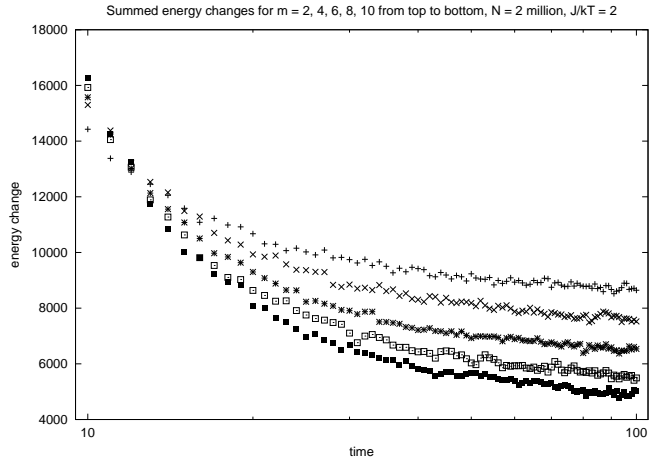}
\end{center}
\caption{Central part of energy curves from Fig. 4 combined with other 
$m$ values, showing $m=2$, 4, 6, 8, 10 from top to bottom, and a relaxation time
increasing with $m$.
Using odd $m$ only the plot (not shown) is similar except that the curve
for $m=1$ is more separated from the others. 
}
\end{figure}

\begin{figure}[hbt]
\begin{center}
\includegraphics[angle=-90,scale=0.35]{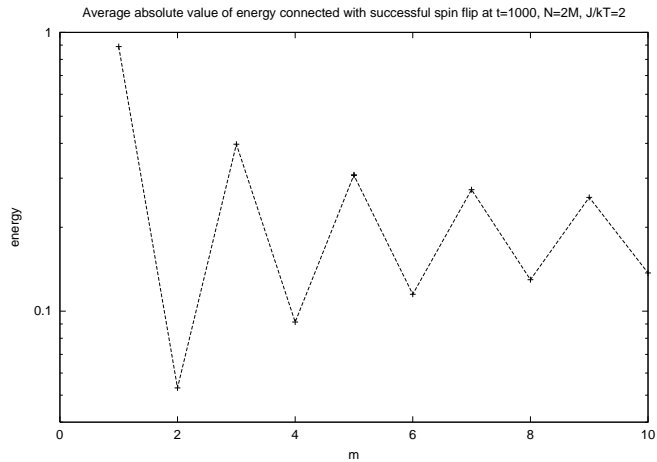}
\end{center}
\caption{Semilogarithmic plot of energy change per spin flip during the last of
1000 iterations (Monte Carlo steps per spin); $N = 2$ million, $J/k_BT = 0.5$.
Every succesful spin flip attempt is taken into account. 
}
\end{figure}

\end{document}